\documentclass[aps,pre,reprint,showpacs,superscriptaddress]{revtex4-1}
\usepackage{graphicx}
\usepackage{dcolumn}
\usepackage{bm}
\usepackage{color}
\usepackage{lineno}
\usepackage{verbatim}
\begin{document}


\title{Particle-in-cell simulations of laser-plasma interactions at solid densities and relativistic intensities: the role of atomic processes}

\author{D. Wu}
\affiliation{State Key Laboratory of High Field Laser Physics, 
Shanghai Institute of Optics and Fine Mechanics, 201800 Shanghai, China}
\affiliation{Helmholtz Institut Jena, D-07743 Jena, Germany}
\author{X. T. He}
\affiliation{Key Laboratory of HEDP of the Ministry of Education, Center for Applied Physics and Technology, 
Peking University, 100871 Beijing, China}
\author{W. Yu}
\affiliation{State Key Laboratory of High Field Laser Physics, 
Shanghai Institute of Optics and Fine Mechanics, 201800 Shanghai, China}
\author{S. Fritzsche}
\affiliation{Helmholtz Institut Jena, D-07743 Jena, Germany}
\affiliation{Theoretisch-Physikalisches Institut, Friedrich-Schiller-University Jena, D-07743 Jena, Germany}

\date{\today}

\begin{abstract}   
   
Direct studies of intense laser-solid interactions is still of great challenges, because of the many coupled physical mechanisms, such as direct laser heating, ionization dynamics, collision among charged particles, and electrostatic or electromagnetic instabilities, to name just a few. Here, we present a full particle-in-cell simulation (PIC) framework, which enables us to calculate laser-solid interactions in a ``first principle'' way, covering almost ``all'' the coupled physical mechanisms. Apart from the mechanisms above, the numerical self-heating of PIC simulations, which usually appears in solid-density plasmas, is also well controlled by the proposed ``layered-density'' method. This method can be easily implemented into the state-of-the-art PIC codes. Especially, the electron heating/acceleration at relativistically intense laser-solid interactions in the presence of large scale pre-formed plasmas is re-investigated by this PIC code. 
Results indicate that collisional damping (even though it is very week) 
can significantly influence the electron heating/acceleration in front of the target. 
Furthermore the Bremsstrahlung radiation will be enhanced by $2\sim3$ times when the solid is dramatically heated and ionized. 
For the considered case, where laser is of intensity $10^{20}\ \text{W}/\text{cm}^2$ and pre-plasma in front of the solid target is of scale-length $10\ \mu\text{m}$, collision damping coupled with ionization dynamics and Bremsstrahlung radiations is shown to lower the ``cut-off'' electron energy by $25\%$. In addition, the resistive electromagnetic fields due to Ohmic-heating also play a non-ignorable role and must be included in realistic laser-solid interactions.       

\end{abstract}


\maketitle

\section{Introduction}
The development of short pulse lasers at relativistic intensities have aroused exciting progress in high energy density physics (HEDP). 
Especially, relativistically intense laser solid interactions are of crucial importance to many great applications, 
such as fast ignition of fusion energy \cite{hedp1,hedp2,hedp3,hedp4,hedp5,hedp6,hedp7,hedp8}, hadron therapy \cite{med1,med2,med3,med4}, 
proton radiography \cite{pg1,pg2,pg3}, high quality ion beam source \cite{ia1,ia2,ia3,ia4,ia5}.  
When an intense laser beam irradiates a solid target, relativistic electrons can be produced in front of the target through the direct-laser-heating/acceleration mechanism.  
These energetic electrons can propagate through the bulk solid and trigger abundant plasma-atomic processes, 
which typically include resistive return current, resistive electric and magnetic fields \cite{rem}, bulk heating and ionization dynamics \cite{bulkionization}, 
Bremsstrahlung X-ray generation \cite{br1,br2,br3} and also ion accelerations \cite{ia1,ia2,ia3,ia4,ia5}. 

In the last decades, there are many worldwide research groups 
focusing on direct-laser-heating of energetic electrons and their transportation in solid target, both experimentally and theoretically. 
These studies can be roughly categorised into two subjects, determined by whether it is the intense laser fields or the solid-density effects that play the dominant roles. 
When an intense laser beam irradiates a bulk solid, it is reflected back by the encountered high density plasmas. 
The two conflicting laser pulses can efficiently accelerate electrons therein 
in front of the target \cite{beg,wilks,lppi1,lppi2,lppi3,lppi4,lppi5,lppi6}. 
This direct-laser-heating/acceleration of energetic electrons is a pure plasma physics process, 
which can be well investigated by the widely used particle-in-cell (PIC) codes. 
Typically, temperature of these energetic electrons can be described by $\textbf{J}\times\textbf{B}$ heating mechanism, which had already  been well formulated by Beg's scaling law \cite{beg} or Wilks' scaling law \cite{wilks}. 
However, when there exists large-scale preformed plasma in front of the solid target, 
the electron heating is beyond predictions of Beg's scaling law or Wilks' scaling law. 
One has to take into account the synergetic effects of both self-generated charge separation electric fields and laser fields, in order to understand the generation of energetic electrons \cite{lppi3,lppi4,lppi5,lppi6}. 
In a recent work \cite{ta1,ta2}, a two-stage electron acceleration model was proposed for relativistically intense laser-solid interactions in the presence of large scale pre-plasmas. The dependence of the electron heating efficiency on both the pre-plasma scale-length and the laser intensity was figured out. 
A scaling law of energetic electrons with $\delta \epsilon \sim(IL_p)^{1/2}$ was obtained, where $I$ is laser intensity and $L_p$ is pre-plasma scale-length. 

In front of the target, it is the intense laser fields that dominate the interactions. 
However except the electromagnetic dynamics, some atomic processes might also play important roles, like field ionization (multi-photon ionization, tunnelling ionization and barrier-suppression ionization) \cite{fieldionization} and 
quantum-electrodynamics (QED) \cite{qed}. In our recent works \cite{fieldionization,qed}, both field ionization and QED models had been established and implemented into the PIC code. However the greatest challenge of PIC simulations is the fact that one has to include the regions where solid-density effects also dominate. The transportation of energetic electrons through the bulk solid could trigger a lot of coupled physical (plasma and atomic) processes. 
To trace these dynamics, several hybrid simulation frameworks were established \cite{hybrid1,hybrid2,hybrid3,hybrid4,hybrid5,hybrid6}. In hybrid simulations, the energetic electrons are treated kinetically using the Vlasov Fokker-Planck approach and the bulk solid is regarded as a resistive fluid. The most recent work \cite{hybrid5,hybrid6} also invoked the Saha Boltzmann model \cite{saha} (or Thomas Fermi model \cite{thomas}) to synchronously update the ionization of the bulk solid. Note for the existing hybrid simulations, the laser-plasma-interactions have not been considered directly. Instead, the energetic electrons are injected with a temperature obeying certain scaling laws. Most of all, the correctness of the hydrodynamic and hybrid methods is based on a very strong assumption, i.e., equilibrium-state assumption. 
Although the hydrodynamic and hybrid methods have acting as workhorses for several decades in the inertial confinement fusion (ICF) research, the time scale of laser-solid interactions at relativistic intensities is much much shorter than that of the ICF studies. The time scale of ICF is on the order of nano-second. The typical time scale of laser-solid interactions at relativistic intensities is on the order of peco-second or even feto-second, therefore the equilibrium-state assumption is no longer correct any more. 

In order to figure out the above dynamics which share significant non-equilibrium features, a very first principle approach should be constructed from the very beginning. Here, in this paper, we have presented a full PIC framework, which enables us to calculate intense laser-solid interactions in a ``first principle'' way, covering almost ``all'' the coupled physical mechanisms. Furthermore, the numerical self-heating of PIC simulations which usually appears in solid-density plasmas is also well controlled by the proposed ``layered-density'' method. This method can be easily implemented into the state-of-the-art PIC codes. 

As an application, the electron heating/acceleration at relativistically intense laser-solid interactions influenced by large scale pre-formed plasmas is re-investigated by this PIC code. 
Results indicate that collisional damping (even though it is very week) 
can significantly influence the electron heating/acceleration in front of the target. 
Furthermore the Bremsstrahlung radiation will be enhanced by $2\sim3$ times when the solid is dramatically heated and ionized. 
For the considered laser of intensity $10^{20}\ \text{W}/\text{cm}^2$ and solid aluminium (Al) target with pre-plasmas scale-length of $10\ \mu\text{m}$, collision damping coupled with ionization dynamics and Bremsstrahlung radiations is shown to lower the ``cut-off'' electron energy by $25\%$. In addition, the resistive electromagnetic fields due to Ohmic-heating also play a non-ignorable role and must be included in realistic laser solid interactions.  

\section{Atomic models and numerical scheme of the PIC framework}
Although the PIC method is a first principle scheme derived from the Vlasov and the coupled Maxwell's Equations, 
it is a tool originally designed to describe plasmas at high temperature and low densities. Typical plasmas of high temperature and low density are fully dominated by the electromagnetic effects. 
However for solid density plasmas (matter), advanced atomic models need to be taken into account. These models should allow to calculate ionizations in a much more natural manner than equilibrium models. These models should also allow to directly
describe the close interactions in the plasmas and thus, accounts for the multi-particle nature of real plasmas. 
In addition, PIC method is a kind of numerical schemes, which usually suffer significant self-heating. In general, it is challenging for a PIC code to simulate extremely dense and low temperature (less than 1 keV) plasmas. This is because the grid size of PIC is restricted by the plasma Debye length, $\lambda_d\sim\sqrt{T_e/n_e}$, in order to avoid the numerical self-heating \cite{numericalheating}. 
Due to the great demand of huge number of grids and/or particles in the PIC simulations of solid-density-plasmas, it is not realistic to perform even with the current fastest 
super-computers. Therefore, in the research fields of relativistically intense laser-solid interactions, to distinguish the PIC approach, one has to solve the facing challenges, both physically and numerically.

\begin{figure}
\includegraphics[width=8.50cm]{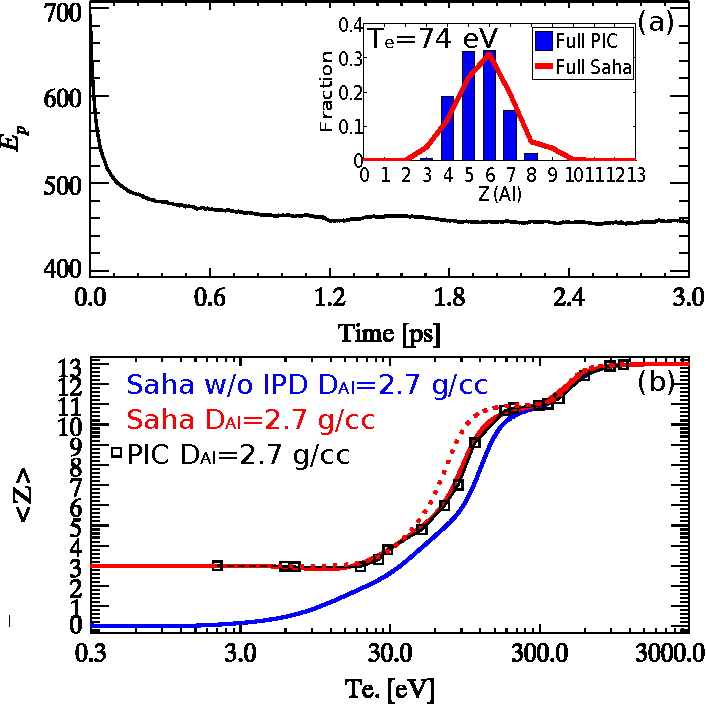}
\caption{\label{fig1} (color online) (a) The total plasma energy (A. U.) within a computational cell as a function of time, with initial plasma temperature $150\ \text{eV}$ and pre-defined charge state $4+$. The red line covered on the inlets are the ionization distributions of Al calculated by Saha-Boltzmann Equation with defined temperature, $T_e=74\ \text{eV}$. 
(b) The averaged ionization degree as a function of temperature, where red and green lines are the results calculated by Saha-Boltzmann Equation, including IPD and excluding IPD, with fixed Al density $2.7\ \text{g}/\text{cm}^3$. 
Solid-red-line is with the SP \cite{ionizationpd1} model of IPD, while dashed red line is with EK \cite{ionizationpd2} model of IPD. 
Black-square-line is picked up from the equilibrium states calculated by our PIC code.}
\end{figure}

\begin{figure}
\includegraphics[width=8.50cm]{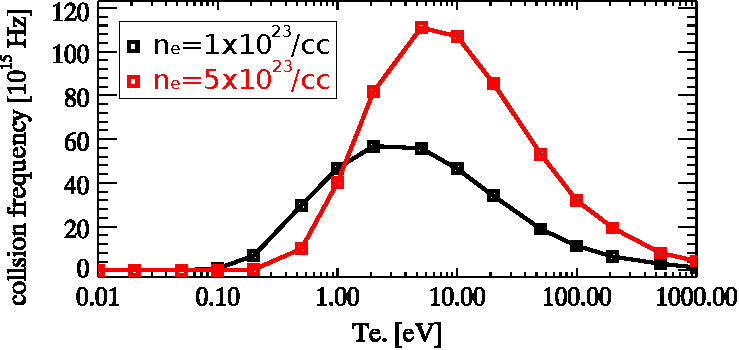}
\caption{\label{fig2} (color online) The plasma collision frequency calculated by the PIC code as functions of temperatures. Here black-square-line refers to fixed plasma density of $10^{23}\ /\text{cm}^3$ and red-square line refers to fixed plasma density of $5\times10^{23}\ /\text{cm}^3$.}
\end{figure}

\subsection{Ionization dynamics}

The main challenge to the ionization dynamics of solid-density plasmas is to incorporate both the matter's response 
to the surrounding plasmas and plasmas' response to the matter. In a recent work \cite{hedp3}, 
we have proposed and analysed a Monte-Carlo approach that can be configured and embedded into the PIC code. 
In this approach, we use a collection of macro-particles to describe a plasma or matter of finite ion density. 
Here, a macro-particle can be regarded as the ensemble of real particles, i.e., a group of particles with ``same'' mass, charge state, position and momentum. The electrons are classified moreover into bound and free ones, where the former are regarded as part of ions or atoms, and the latter are isolated as the surrounding plasmas. Here, both impact (collision) ionization (CI) \cite{cionization} and 
electron-ion recombination (RE) \cite{recombination} are taken into account. 
Furthermore, the ionization potential depression (IPD) \cite{ionizationpd1,ionizationpd2} by the surrounding plasmas is also taken into consideration.

When compared with Saha Boltzmann or Thomas Fermi models, which are applied in the literature for plasmas near thermal equilibrium, 
the temporal relaxation of ionization dynamics can also be simulated by the recently proposed model. Here as a benchmark, the ionization dynamics of an Al bulk (with density $2.7\ \text{g}/\text{cm}^3$) is calculated with our PIC code. We consider only a few computational cells, connected by periodic boundary conditions, with each cell contains $200$ ion macro-particles and $200$ electron macro-particles initially. Fig.\ \ref{fig1} (a) shows the total plasma energy (A. U.) within a computational cell as a function of time, where the initial Al charge state is assumed to be $4+$, and the initial free electron temperature is set to $150\ \text{eV}$. Following the energy history, at initial time, the CI rate of Al is larger than RE. The former one would reduce the plasma energy and increase the averaged ionization degree as a function of time. After $6$ ps relaxation, 
the averaged ionization degree is $\bar{Z}=5.82$ with $T_e=74\ \text{eV}$. In Fig.\ \ref{fig1} (a), the ionization distributions calculated by 
Saha-Boltzmann Equation is also present in the red curves covered on the inlets, 
showing good consistence with the PIC calculations. Following the same routine, the dependence of averaged ionization degree on thermal equilibrium temperatures covering a large variation is obtained by the PIC code, as shown in black-square-lines in Fig.\ \ref{fig4} (b), also showing good consistence with results from Saha-Boltzmann Equation.  

\subsection{Collision with Bremsstrahlung corrections}

It is well known in plasma physics that electron-electron, electron-ion and ion-ion scatterings can be described by means of the Monte-Carlo binary collision model, thanks to the pioneering works of Takizuka \cite{pic_collision1}, Nanbu \cite{pic_collision2} and Sentoku \cite{pic_collision3}. 
Within these PIC calculations, three steps are made iteratively: i) pair of particles are selected randomly in the cell, i.e. either electron-electron, electron-ion or ion-ion pairs; ii) for these pair of particles, the binary collisions are associated with changes in the velocity of the particles within the time interval $\delta t$ and which are calculated; iii) and then the velocity of each particle is replaced by the newly calculated one. The collision frequency of fully ionized plasmas between charged particles, 
used in these PIC calculations, is $\nu={8\sqrt{2\pi} e^4Z_a^2 Z_b^2 n_{\min}}\ln{(\Lambda_{\text{f}})/(3m_e^2\beta^3)}$, 
where $Z_a$ and $Z_b$ are charge state of colliding particles, $n_{\min}$ is the minimal density of the two species $a$ and $b$, 
and $\beta$ is the relative velocity between the two colliding particles.  
The Coulomb logarithm, $\ln{(\Lambda_{\text{f}})}$, is usually defined as $L\equiv\ln(\lambda_{\text{D}}/b)$, 
where the Debye length, $\lambda_{\text{D}}$, is a dynamic value changing as $\lambda_{\text{D}}=\sqrt{(T_e/4\pi n_e)(1+\beta^2/v^2_{\text{th}})}$, 
where $T_e$ and $v_{\text{th}}$ are the temperature and thermal velocity of background electrons. Parameter $b$ is the distance of closest approach between the two charges. In classical scattering, we have $b=Z_a Z_b e^2/m_e\beta^2$. 
This condition is not satisfied in the relativistic case, so that the scattering must be treated quantum-mechanically using the Born approximation. In this case, i.e., $e^2/\hbar \beta\ll1$, the Coulomb logarithm is then expressed as $L=\ln{(\lambda_{\text{D}}\gamma \beta/\hbar)}$, which is the ratio of the Debye length and the de Broglie wave length. This definition of Coulomb logarithm works well for low-density and high-temperature plasmas. 
However, for plasmas of solid-density and at low temperatures, as $b$ will be larger than $\lambda_{\text{D}}$, 
the Coulomb logarithm expression will become negative. Existing works of Takizuka, Nanbu and Sentoku do not address this issue of negative Coulomb logarithm, as the collision models are initially proposed for high temperature plasmas. 

\begin{figure}
\includegraphics[width=8.50cm]{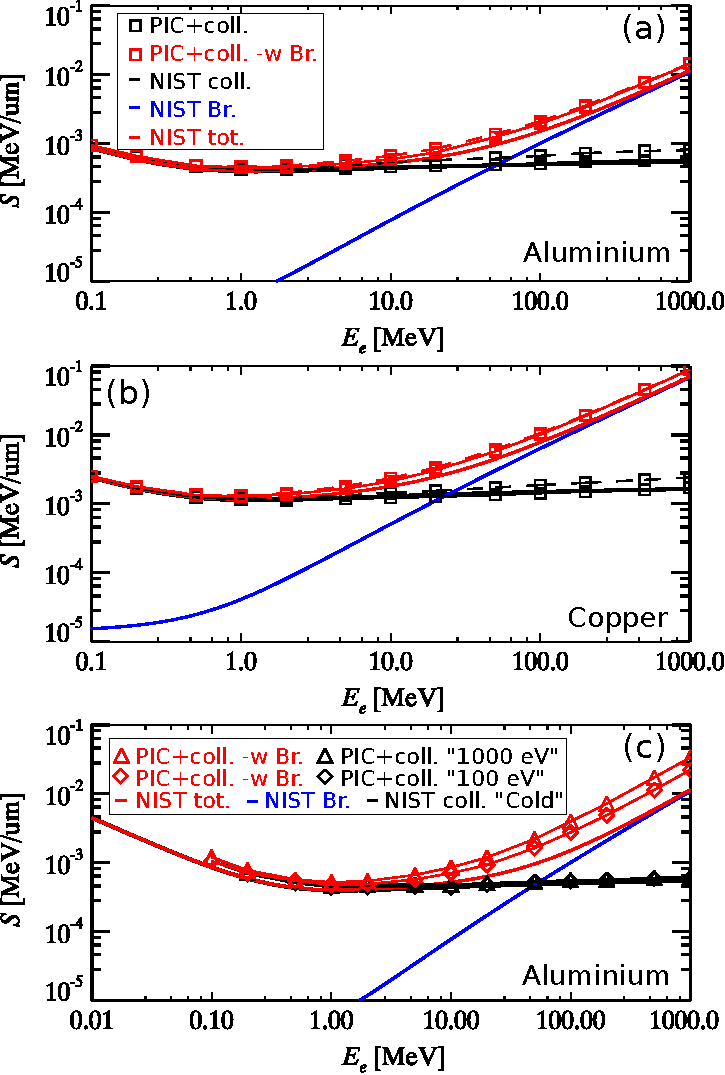}
\caption{\label{fig3} (color online) Stopping power of different materials, (a) for Al and (b) for Cu, as a function of projected electron kinetic energy. 
Results from our PIC simulations, at low temperature limit, are compared with that from the NIST database. 
Solid-black-line is the collisional stopping power ($S_{\text{c}}$), solid-blue-line is the radiation stopping power ($S_{\text{r}}$) 
and solid-red-line is the total stopping power, with $S_{\text{t}}=S_{\text{c}}+S_{\text{r}}$ from the NIST database. 
Black-square-line is the collisional stopping power calculated by PIC code, and red-square-line is the total stopping power calculated by PIC code, 
where dashed-lines represent the one excluding density effect $\delta/2$. 
(c) represent the stopping power of Al as a function of projected electron kinetic energy at different temperatures.}
\end{figure}

We here obtain a general Coulomb logarithm by considering the scattering of charged particles by sheathed Coulomb force, $\exp{(-r/\lambda_{\text{D}})}/r$. Rigorous calculation results in the expression of Coulomb logarithm as $L=\ln[(1+\eta)/\eta]$ (Appendix A), where $\eta=b/\lambda_{\text{D}}$. 
This expression of Coulomb logarithm will converge to $L=\ln{(\lambda_{\text{D}}/b)}$ when $b\ll\lambda_{\text{D}}$ for high temperature plasmas. In addition, the collision in the degenerate regime is also taken into account, with frequency $\nu=(4m_ee^4/3\pi \hbar^3)L$, 
where $L$ has the same definition of $L=\ln[(1+\eta)/\eta]$. When $\beta$ is smaller than the value $n^{1/3}_{\max}\hbar/m_ec$, where $n_{\max}$ is the maximal density between species $a$ and $b$, $\nu=(4m_ee^4/3\pi \hbar^3)L$ is used instead. 
For given plasma density, Fig.\ \ref{fig2} shows the PIC-code-calculated collision frequencies as functions of temperatures. 
Plasmas with fixed density of $1\times10^{23}\ /\text{cm}^3$ is shown in black-square-line and 
fixed density of $5\times10^{23}\ /\text{cm}^3$ is shown in red-square-line. 
The collision frequency is calculated by averaging over $1000$ pairs of colliding particles. For high temperature, $T_e\gg10\ \text{eV}$, the collision frequency nicely converges to the Spitzer model \cite{pic_collision1,pic_collision2,pic_collision3}, i.e., $\nu\sim T_e^{-1.5}$, which decreases rapidly with the raising of temperatures. While at low temperature limit, 
when $\lambda_{\text{D}}$ is extremely small, the collective behaviour is significantly depressed. Instead, the charged particles trend to interact with each other more like rigid-ideal-gas, where the collision frequency is increasing with the increase of temperatures.  

The above collision model works well for fully ionized plasmas. However in laser-solid interactions, the inner part of the bulk target is usually of partially ionized, therefore the contribution of bound electrons must be taken into account. In a recent work \cite{hedp4}, we have studied the ion stopping in warm dense matter (or/and partially ionized plasma), where both the contribution of bound and free electrons are included by modifying the ion-electron collision frequency as,
\begin{equation}
\label{scattering}
\nu_{\text{i-e}}=\frac{8\sqrt{2\pi} e^4Z^2_bZn_{i}}{3m_e^2\beta^3}[\ln{(\Lambda_{\text{f}})}+\frac{A-Z}{Z}\ln{(\Lambda_{\text{b}})}],
\end{equation}   
where $\ln{(\Lambda_{\text{b}})}\equiv\ln{|{2\gamma^2 m_e\beta^2}/{\bar{I}_{A}(Z)}|}-\beta^2-{C_{\text{K}}}/{A}-{\delta}/{2}$, $I_{A}(z)$ is the effective ionization potential, $\delta/2$ is the density effect contribution
and ${(A-Z)}/{Z}$ ($A$ is the atomic number, $A=13$ for Al, and $Z$ is the ionization state) defines the ratio of bound electrons' contributions. 
For a fully ionized plasmas, $Z\to A$, the collision frequency between ions and electrons in Eq.\ (\ref{scattering}) converges to $\nu_{\text{i-e}}\sim[{8\sqrt{2\pi} Z^2_be^4 Z n_{i}}/{3m_e^2\beta^3}]\ln{(\Lambda_{\text{f}})}$. 
For neutral atoms, $Z\to 0$, in contrast, the frequency in Eq.\ (\ref{scattering}) is $\nu_{\text{i-e}}\sim[{8\sqrt{2\pi} Z^2_be^4 A n_{i}}/{3m_e^2\beta^3}]\ln{(\Lambda_{\text{b}})}$. If the projectile is electron, the value of $\ln(\Lambda_{\text{b}})$ must be estimated in the center-of-mass frame, and the expression becomes $\ln{(\Lambda_{\text{b}})}\equiv\ln{|(\gamma-1)\sqrt{(\gamma+1)/2}m_ec^2/{\bar{I}_{A}(Z)}|}-\beta^2/2-\delta/2$. 

\begin{table}
\caption{\label{table1} Coulomb logarithm and $\delta/2$ as a function of energy of projected electrons for solid Al and Cu 
at low temperature limit, where $\ln(\Lambda_{\text{b}})$ are calculated in the PIC code by averaging over $1000$ projected electrons and values of $\delta/2$ are obtained from the NIST database.}
\begin{ruledtabular}
\begin{tabular}{ l  l  l  l  l  l  l  l }
MeV & 1.0 & 5.0 & 10.0 & 50.0 & 100.0 & 500.0 & 1000.\\
$\ln(\Lambda_{\text{b}})_{\text{Al}}$ & 8.59 & 10.70 & 11.68 & 14.06 & 15.09 & 17.50 & 18.55 \\ 
$\ln(\Lambda_{\text{b}})_{\text{Cu}}$ & 7.76 & 9.86  & 10.86 & 13.23 & 14.26 & 16.67 & 17.72 \\ 
$(\delta/2)_{\text{Al}}$              & 0.33 & 1.43  & 2.38  & 5.07  & 6.36  & 9.53  & 10.92 \\
$(\delta/2)_{\text{Cu}}$              & 0.58 & 1.85  & 2.66  & 5.05  & 6.29  & 9.37  & 10.74 \\
\end{tabular}
\end{ruledtabular}
\end{table}

\begin{figure}
\includegraphics[width=8.50cm]{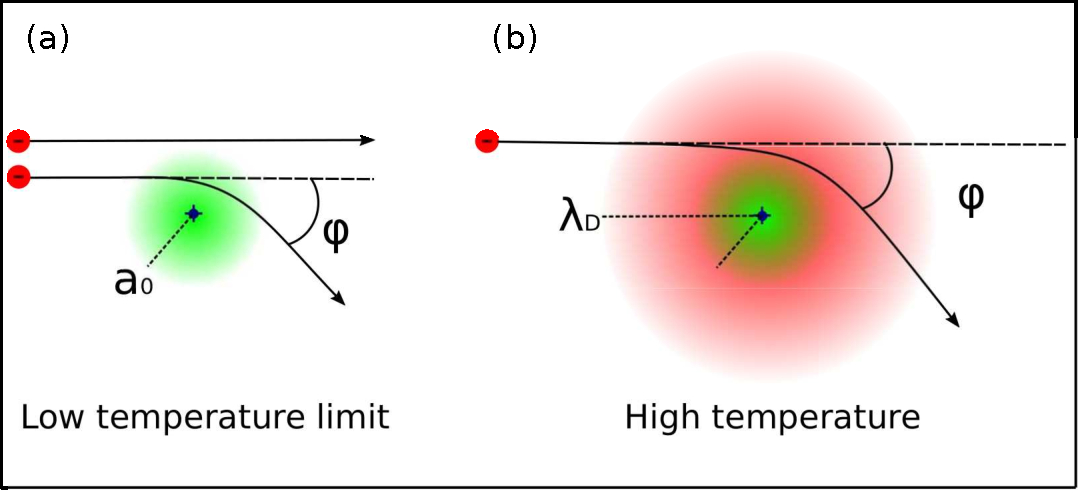}
\caption{\label{fig4} (color online) Schematic of charged particle collision. For neutral atom, when all electrons are bound at the nuclei with the radius on the order of Bohr unit $a_0$, only projectile that could penetrate through the electron can be deflected by the Coulomb force of the nuclei. When temperature is high, some of the bound electrons are ionized and form plasmas around the nuclei, projectile with a collision distance $b$ smaller than $\lambda_{\text{D}}$ (usually $\lambda_{\text{D}}$ is much larger than $a_0$) can also be deflected by the Coulomb force.}
\end{figure}

As a benchmark of our collision model, Fig.\ \ref{fig3} shows the variation of stopping power as functions of energy, when energetic 
electrons transport through a bulk solid. 
Solid-black-line represents the collisional stopping power ($S_{\text{c}}$) obtained from 
the National Institute of Standard and Technology (NIST) \cite{NIST} database. 
Results obtained from our PIC simulations (black-square-lines) are also listed to compare with that from the NIST database. 
Simulation results from the PIC code nicely reproduce the collisional stopping powers as obtained from the NIST database, 
for both Al as shown in (b) and copper (Cu) as shown in (b). 
For the stopping power calculation of energetic electrons, the density effect, i.e., $\delta/2$ term, plays an important role. 
When excluding this effect, as dashed-black-square-lines show, the stopping power is significantly larger than the values from the NIST database. 
For the corrections $\delta/2$, which is due to the electron density of the target, no simple relationship is available between the
magnitude and atomic number of the stopping medium. Fortunately, it have already been tabulated for all elemental targets \cite{NIST,delta2}. 
In Table.\ \ref{table1}, we have organized $\ln(\Lambda_{b})$ and $\delta/2$ as functions of electron kinetic energies, 
where $\ln(\Lambda_{\text{b}})$ are calculated with the PIC code by averaging over $1000$ projected electrons and values of density effects are obtained from the NIST database. It is shown that the value of density effect is increasing with the raising of projected electron energy. It is comparable to that of $\ln(\Lambda_{b})$ when the projected electron energy is high, especially when the kinetic energy is of $E_k\gg10\ \text{MeV}$.

\begin{figure}
\includegraphics[width=8.50cm]{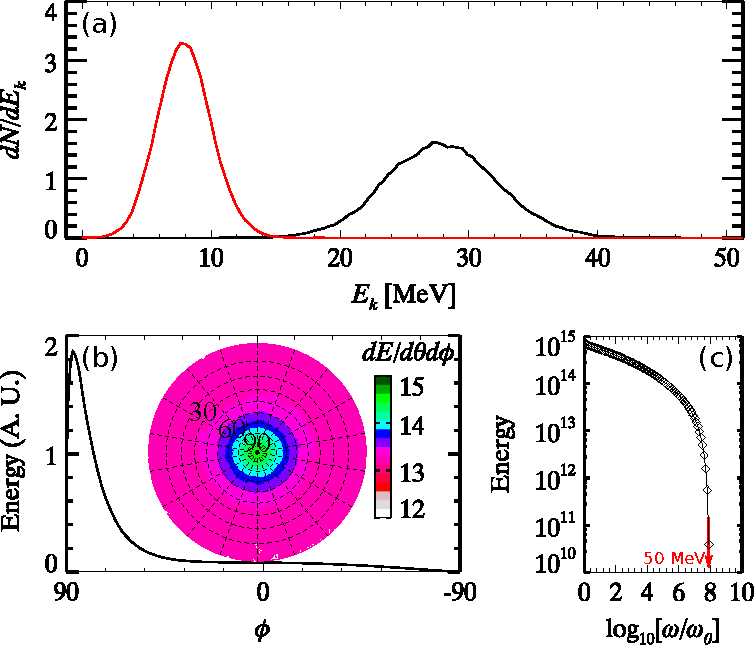}
\caption{\label{fig5} (color online) Comparison of PIC simulations when including and excluding Bremsstrahlung radiation correction. Initially, a mono-energetic electron beam of $E=50\ \text{MeV}$ is launched into a bulk Al. The finial energy spectrum after $150\ \text{ps}$ is shown in (a), where red-line is the case including Bremsstrahlung and black-line is the one excluding Bremsstrahlung. (b) is the angular distribution of emitted photons due to Bremsstrahlung radiation. See text for the explanation of coordinate set-up. 
(c) is the frequency spectra of emitted photons due to Bremsstrahlung radiation, 
where we have plotted $\int_{\hbar\omega_{\text{k}}}^{\infty}[dE/d(\hbar\omega)]d(\hbar\omega)$ as function of cut-off frequency $\omega_{\text{k}}$. 
Note $\hbar\omega_{0}=1.24$ eV, corresponding to the energy of a photon with wavelength $1\ \mu\text{m}$.}
\end{figure}

\begin{figure}
\includegraphics[width=8.50cm]{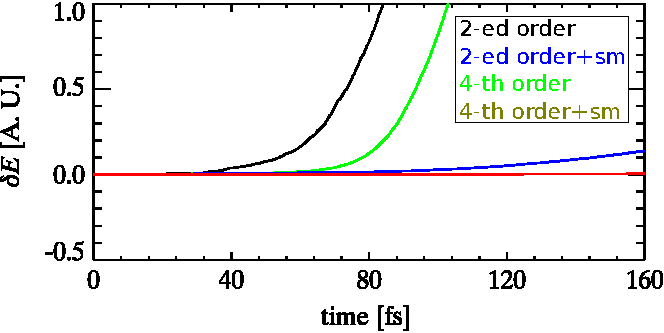}
\caption{\label{fig6} (color online) The value of self-heating as function of simulation time. Plasma is of density $100n_c$ ($n_c$ is the corresponding critical density for electromagnetic wave of wavelength $1\ \mu\text{m}$), and plasma temperature is $T_e=10\ \text{eV}$, 
the simulation grid size is $0.02\ \mu\text{m}$ and $100$ electrons are filled into a computational cell. Different coloured lines represent different combinations of numerical schemes, $4$-th/$2$-ed order and with/without current smoothing.}
\end{figure}

When kinetic energy of projected electrons is high, for example $E_k\gg10\ \text{MeV}$, the radiation stopping also becomes non-ignorable. This is because, when charged particles collide, they will accelerate in each other's electric field and as a result, radiating electromagnetic waves. Generally, the total energy radiated in this collision is given for the instantaneous radiated power by an accelerated charge, $P\sim\dot{\beta}^2Z^2$, integrated over the duration time of collision, $\tau$. For an energetic electron propagates through a target, following Jackson \cite{jackson}, 
we can obtain the energy radiated per unit length per unit frequency as,
\begin{equation}
\label{dEdldw}
\frac{d^2E}{dld(\hbar\omega)}=\frac{16}{3}\alpha r^2_e n A^2\ln|\frac{2\gamma\gamma^{\prime}m_ec^2}{\hbar\omega}|
\end{equation}  
where $n$ is the ion density of the target, $A$ is the atomic number of the target material ($A=13$ for Al), 
$\alpha=e^2/\hbar c$ is fine structure constant, $r_e=e^2/m_ec^2$ is classical electron radius and $\gamma^{\prime}=\gamma-\hbar\omega$ is the relativistic factor of the electron after the photon has been emitted. 
For energetic electrons, the radiation of energetic electrons is emitted mainly in the forward direction. The average angle between the directions of motion of the electron and the emitted light is of the order $\sim1/\gamma$. Therefore in PIC simulations, the angular distribution of emitted photons can be approximated as
\begin{equation}
\frac{dE}{d\Omega d(\hbar\omega)}=\frac{4}{3\pi}\delta(1-\frac{\bm{p}}{|\bm{p}|})\alpha cr^2_e nA^2\ln|\frac{2\gamma\gamma^{\prime}m_ec^2}{\hbar\omega}| dt,
\end{equation} 
where a delta-function approximation is used to describe the direction of photon emissions. 

Following the definition of collision stopping power, the radiation stopping power should have the following form by integrating Eq.\ (\ref{dEdldw}) with $\hbar\omega$, which is from $0$ to $\gamma m_ec^2$,  
\begin{equation}
\label{dEdl}
S_{\text{r}}\equiv\frac{dE}{dl}=\frac{16}{3}\gamma m_ec^2\alpha r^2_e n A^2\ln(\Lambda).
\end{equation} 
In Eq.\ (\ref{dEdl}), we need to account for the screening of nuclear potential by surrounding electrons at the nuclei when the collisions are distant. 
Similar with the approach applied for the Coulomb logarithm calculation (Appendix A), 
when impact parameter is larger than a particular value, the potential is artificially set to be zero. At low temperature limit, when all electrons are bound at the nuclei, the ``Thomas-Fermi'' potential is an approximation to the screened nuclear potential. 
It can be approximated as $\phi=(Ze/r)\exp(-r/a)$, with the characteristic length $a=1.4a_0A^{-1/3}$, where $a_0$ is the Bohr unit. 
This kind of screening will reduce the power radiated, because it essentially lowers the maximal effective impact parameter to $\sim a$. 
For relativistic collisions, we replace the characteristic maximum impact parameter $2\gamma\gamma^{\prime}c/\omega$ with $a$ if $a$ is smaller. It will be if $(\omega/2\gamma^2c)(1.4a_0/A^{1/3})<1$. This inequality will apply over the entire frequency range up to the maximum possible photon energy $\hbar\omega=\gamma m_ec^2$ if the incident energy satisfies $(m_ec^2/2\gamma\hbar c)(1.4a_0/A^{1/3})<1$. 
For a particular material, such as Al, this inequality means when energy of the colliding electron is of $E_k>20.3\ \text{MeV}$, 
the screening is important and $\ln(\Lambda)$ in Eq.\ (\ref{dEdl}) can be re-written as constant $\ln(\Lambda)\equiv\ln|233/A^{1/3}|$. While for Cu, the threshold of screening is $E_k>15.4\ \text{MeV}$. If electron energy is higher than the threshold, 
radiation stopping power, as represented by Eq.\ (\ref{dEdl}), 
is a linear function of energy. This kind of behaviour is also well confirmed by the solid-blue-line, picked from the NIST database, as shown in Fig.\ \ref{fig3}.

When temperature is high, some of the bound electrons are ionized and form plasmas surrounding the nuclei. As schematically shown in Fig.\ \ref{fig4}, 
this will increase the maximal effective impact parameter from $\sim a$ to $\lambda_{\text{D}}$, here $\lambda_{\text{D}}=\sqrt{T_e/4\pi n_e}$ is the Debye length of plasmas. When including ionization effect, the radiation stopping power, which is originally shown in Eq.\ (\ref{dEdl}), can be re-written as
\begin{equation}
\label{dEdlmd}
S_{\text{r}}\equiv\frac{16}{3}\gamma m_ec^2\alpha r^2_e n Z^2[\frac{A^2}{Z^2}L_a+L_{\text{D}}],
\end{equation}
where $L_a=\ln|{a m_ec}/{\hbar}|$ and $L_{\text{D}}=\ln|{\lambda_{\text{D}}}/{a}|$. This updated radiation stopping power can converge to neutral atom limit when $Z\rightarrow0$ and also to purely plasma cases when $Z\rightarrow A$, note that $L_a+L_{\text{D}}=\ln|\lambda_{\text{D}}m_ec/\hbar|$.

In PIC simulations, as the average angle between photon and electron can be handled by a delta-function approximation, the Bremsstrahlung radiation do not further change the deflection of the electron. This approximation will significantly simplify the implementation of Bremsstrahlung correction into the binary collision models. In the binary collision model, right after the second step of calculation cycles, the electron energy is updated by including the Bremsstrahlung correction, i.e., $\gamma_{\text{Br}}=\gamma-\delta\gamma$ with $\delta\gamma m_ec^2=c\delta t S_{\text{r}}$. 
The electron momentum is also updated with $\bm{p}_{\text{Br}}=\sqrt{(\gamma_{\text{Br}}-1)/(\gamma-1)}\bm{p}$, where $\delta t$ is the time step of PIC simulations, and $n$ in Eq.\ (\ref{dEdlmd}) should also be replaced by the minimal density $n_{\min}$ between injected electrons and target ions. 

When the Bremsstrahlung radiation correction is included, red-square-lines in Fig.\ \ref{fig3} (a) and (b) show the total stopping power (including both radiation and collision) as functions of projected electron energy, where (a) is for Al and and (b) is for Cu. 
Our stopping power values calculated by the PIC code nicely reproduce that from the NIST datebase. Note that datas from the NIST databases are obtained at the neutral atom limit. When temperature is high, some of the bound electrons are ionized and form surrounding plasmas, the corresponding radiation stopping power should also be increased accordingly. As shown in Fig.\ \ref{fig3} (c), the total stopping powers calculated by the PIC code at different temperatures, $T_e=100\ \text{eV}$ with $Z=8.0$ (red-diamond-line) and $T_e=1000\ \text{eV}$ (red-triangle-line) with $Z=13$, are presented. 
As expected, the radiation stopping power is increased by $2\sim3$ times when the temperature (and ionization) of target is increased to hundreds of eV.   

In order to detail the comparison between collision model with and without the Bremsstrahlung radiation correction, the ``EM-field mode'' in PIC simulations is turned off. Initially, a mono-energetic electron beam of $E_k=50\ \text{MeV}$ is launched into a bulk Al. After $150$ ps, the finial energy spectrum with and without Bremsstrahlung radiation correction are shown in Fig.\ \ref{fig5} (a). The red-line is the case including Bremsstrahlung and black-line is the one excluding Bremsstrahlung. We can see, as expected, the peak energy of electron beam is $8\ \text{MeV}$ (with Bremsstrahlung radiation correction) v.s. $28\ \text{MeV}$ (without Bremsstrahlung radiation correction). The energy spread is also significantly contracted by the Bremsstrahlung radiation.
 
The angular distribution of emitted photons due to Bremsstrahlung radiation is presented in Fig.\ \ref{fig5} (b). Here the definition of angle is similar to the longitude and latitude system on a map of Earth. Here the longitude angle $\theta$ spans from $-180$ to $180$, 
which is defined as azimuthal angle between the X-axis and the transverse momentum of the photon. The latitude angle $\phi$ spans from $-90$ to $90$, which is defined as the angle between the laser optical Y-axis and the photon propagation direction. We can see that the radiation of photons is almost of forward direction, although a slight deflection from the Z-axis of $3\sim5$ degree is observed. In Fig.\ \ref{fig5} (c), we also present the frequency spectrum of emitted photons, 
where we have plotted $\int_{\hbar\omega_{\text{k}}}^{\infty}[dE/d(\hbar\omega)]d(\hbar\omega)$ as function of cut-off frequency $\omega_{\text{k}}$. 
Note the cut-off energy of the radiated photons is exactly equal to the maximum electron energy, i.e., $50\ \text{MeV}$.
The cut-off frequency is as high as  $\sim0.4\times10^{8}\hbar\omega_{0}$, where $\hbar\omega_{0}=1.24\ \text{eV}$, which is the energy of a photon with wavelength $1\ \mu\text{m}$.
    
\subsection{``Layered density'' method PIC}

It is well known that PIC codes are prone to a phenomenon known as self-heating. 
In general, the grid size of PIC is restricted by the plasma Debye length $\lambda_{\text{D}}\sim\sqrt{T_e/n_e}$, to avoid the numerical self-heating \cite{numericalheating}. However the analysis presented there concentrates primarily on the case in which particle forces are assigned to the nearest-neighbour grid points. Here we refer this method as the $1$-st order scheme.
Recently, high-order explicit electromagnetic fields solver and smoother particle shape functions have been implemented into PIC codes. Significant advantages over $1$-st order scheme have been reported \cite{highorder}, and the restriction of grid size in PIC simulations is also increased from plasma Debye length $\lambda_{\text{D}}$ to skin depth $l\sim\sqrt{m_ec^2/n_e}$. 

\begin{figure}
\includegraphics[width=8.50cm]{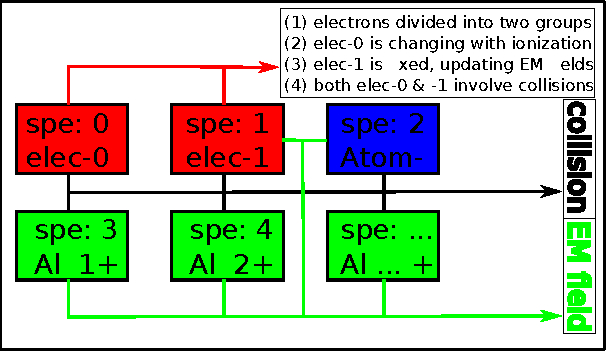}
\caption{\label{fig7} (color online) The schematic of ``layered density'' method. Here ``layered density'' means electrons are divided into two groups, electron-0 and electron-1. During the PIC simulation, electron-0 is a changing, which updates following the ionization dynamics. When calculating electromagnetic fields, only electron-1 is involved in. For collisions, both electron-0 and electron-1 are involved in.}
\end{figure} 

\begin{figure}
\includegraphics[width=8.50cm]{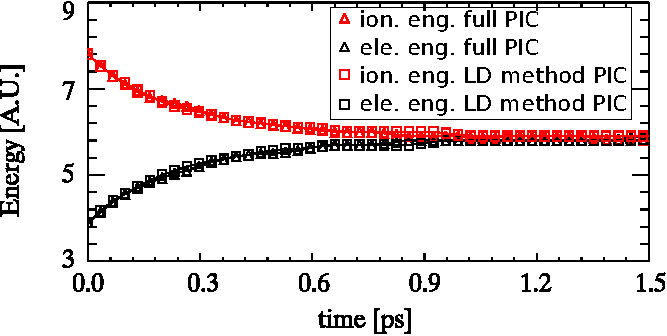}
\caption{\label{fig8} (color online) Thermal equilibrium benchmark of the ``layered density'' (LD) method. 
Electron and ion kinetic energy as function of time. Initial plasma density is set to be $100n_c$, initial electron temperature is $50\ \text{eV}$ and initial proton temperature is $100\ \text{eV}$. For the LD method, electrons are divided into two groups, and the density of each group is $50n_c$. In PIC simulations, these two groups of electron are treated as \textit{different} species.}
\end{figure}

\begin{figure}
\includegraphics[width=8.50cm]{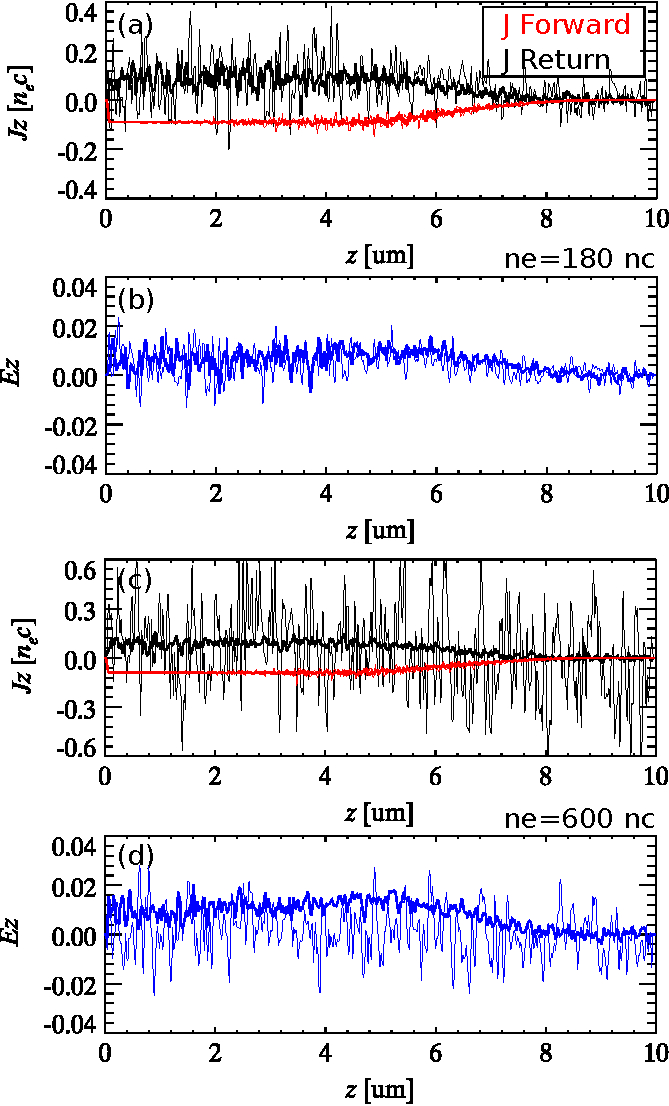}
\caption{\label{fig9} (color online) (a) and (c) The current density distribution, $\bm{J}$, of forward-propagating fast-electrons (red line) and returning background-electrons (black line), when a fast electron beam of $1$ MeV with density $0.1n_c$ is launched into uniform plasmas. (b) and (d) The resistive electric fields generated by the launched electron beam. Here background plasma density in (a) and (b) is of $180n_c$ ($n_c$ is the corresponding critical density for electromagnetic wave of wavelength $1\ \mu\text{m}$) and temperature is of $T_e=10\ \text{eV}$. In (c) and (d), plasma density of $600n_c$ is used. Thick lines are the results calculated by ``layered density'' methods, and thin lines are the one obtained from full-PIC method.}
\end{figure} 

In Fig.\ \ref{fig6}, we have presented the value of self-heating as function of simulation time. In these simulations, 
the plasma density is $100n_c$. Here $n_c$ is the corresponding critical density for electromagnetic wave of wavelength $1\ \mu\text{m}$. 
The plasma temperature is of $T_e=10\ \text{eV}$. The plasma Debye length is $\lambda_{\text{D}}=4\times10^{-6}\ \mu\text{m}$ and skin depth is $l=0.024\ \mu\text{m}$. The simulation grid size is $0.02\ \mu\text{m}$, which is smaller than skin depth.
We fill $100$ electrons into each computational cell. Different coloured lines represent the combinations of different numerical schemes. 
In Fig.\ \ref{fig6}, larger and smoother particle shape functions coupled with multi-points electromagnetic field solvers are regarded as high order schemes. It is clearly demonstrated that $4$-th order numerical scheme coupled with current smoothing technique shows significant advantage over others. 
Therefore, in our following simulations, this combination is regarded as the default set-up. 

The combination of $4$-th order numerical scheme coupled with current smoothing technique is a useful approach to avoid significant numerical self-heating. However if plasma density is further increased, it is still a great challenge for the present PIC codes. 
For example the electron density can be as high as $10^{24}\ /\text{cm}^3$ in solid metals, or even as high as $10^{25}\ /\text{cm}^3$ in the compressed D-T core which exists in fast-ignition inertial confinement fusion research.
It is now a well accepted fact that the numerical self-heating arises from the high density background plasmas. For extremely high density plasmas, it is the collision effects that dominant, while the electromagnetic effects trend to be significantly suppressed. 
If one turn off the electromagnetic field solver for the high density background electrons, the self-heating can be definitely avoided. However, by doing so, one also lost some important physics, like generation of return current and resistive electric and magnetic fields. 

Here we suggest a ``layered density'' method, which can well deal with plasmas with extremely high densities. This method is not a rigorous numerical scheme, but an empirical method. A schematic structure is shown in Fig.\ \ref{fig7}. In this method, we divide the high density background electrons into two groups, regarded as ele-$0$ and ele-$1$. In PIC simulations, although both ele-$0$ and ele-$1$ have the same mass and charge, 
they are treated as \textit{different} particle species. 
Usually density $n_{e0}$ of ele-$0$ is close to the original background density $n_e$, 
while density of ele-$1$ is $n_{e1}=n_e-n_{e0}$. In PIC simulations, the movement of charged particles generates a distribution of current density, 
and this current density will in turn updates the electromagnetic fields. In the ``layered density'' method, the contribution to the current density from ele-$0$ is turned off, while only ele-$1$'s contribution is reserved.  
The variation of $n_{e0}$ is restricted to ionization dynamics, while the variation of $n_{e1}$ is due to the actions of electromagnetic fields. 
Both ele-$0$ and ele-$1$ involve in the collision effects. 
Although, this ``layered density'' method can avoid numerical self-heating, 
whether this kind of set-up is applicable or not still demands rigorous benchmarks: i) thermal equilibrium benchmark; ii) Ohmic return current and resistive electric field or magnetic field benchmark. 

In Fig.\ \ref{fig8}, thermal equilibrium benchmark of the ``layered density'' method is demonstrated. 
In this benchmark, initial plasma density is set to be $100n_c$, 
initial electron temperature is $50\ \text{eV}$ and initial proton temperature is $100\ \text{eV}$. 
The black-triangle and red-triangle lines represent the 
relaxation process of electrons and protons with initially different temperatures. For the ``layered density'' method, 
electrons are divided into two groups, 
and the density of each group is $50n_c$. In PIC simulations, these two groups of electron are treated as \textit{different} species. 
The black-square and red-square lines represent the one calculated by ``layered density'' method PIC. 
When comparing with results obtained by the full PIC, we do not find any significant differences. This is because, the collision frequency between charged particles is a linear function of density, i.e., $\nu\sim n_e$. Therefore, a linear decomposition of electrons into different sub-groups do not affect the whole collision dynamics.  

For extremely high density plasmas, the electromagnetic effects are significantly suppressed. However, if one turn off the electromagnetic field solver for the high density background electrons, some important physics, like generation of return current and resistive electric or/and magnetic field, are lost. 
In the 	``layered density'' method, the high density background electrons are divided into two groups, ele-$0$ and ele-$1$, and only the later one are set up to update the electromagnetic fields. As a benchmark of return current and resistive electromagnetic fields, we consider a fast electron beam of $1$ MeV with density $0.1n_c$ launching into uniform plasmas. In Fig.\ \ref{fig9} (a) and (b), the density and temperature of the uniform plasmas are set to $180n_c$ ($n_c$ is the corresponding critical density for electromagnetic wave of wavelength $1\ \mu\text{m}$) and $10\ \text{eV}$. 
The corresponding skin depth is $0.021\ \mu\text{m}$, and grid size in PIC simulation is set to $0.02\ \mu\text{m}$. 
As shown in Fig.\ \ref{fig9} (a), thin-red-line is the current density $J_{\text{fw}}$ of launched fast electrons calculated by full PIC, 
and black line is the Ohmic return current $J_{\text{rt}}$. We can see that the total current is \textit{almost} zero, because the $J_{\text{fw}}$ is almost compensated by $J_{\text{rt}}$. The thick-red and -black-lines are the one calculated by ``layered density'' method PIC, where density of ele-$0$ is $130n_c$ and density of ele-$1$ is $50n_c$. When comparing the results obtained by two different methods, we do not find any significant differences, except that the numerical noises calculated by ``layered density'' method PIC is significantly depressed. 
The corresponding resistive electric field is shown in Fig.\ \ref{fig9} (b), where thin-blue-line represents the one calculated by full PIC and 
thick-blue-line is the one by ``layered density'' method PIC. As for the resistive electric fields, except that the numerical noises calculated by ``layered density'' method PIC is relatively small, we also do not find any significant differences between them. When increasing the uniform background plasma density from $180n_c$ to $600n_c$ and keeping other parameters the same, as shown in Fig.\ \ref{fig9} (c) and (d), the results obtained from full PIC are fully ``swallowed'' by numerical noises. In contrast, the results calculated by the ``layered density'' method PIC are proved to be stable.
 
It seems that the decomposition of high density background plasmas is an arbitrary approach, however one still need to obey some restricted rules.
When a fast electron beam launches into high density plasmas, the Ohmic return current will increase with time, asymptotically approaching steady-state ``Spitzer-limit current'' \cite{rc} given by Ohm's law $\bm{J}=\sigma \bm{E}$. 
The variation of return current density with time can be obtained by seeking a time-dependent solution to the Drude model \cite{rc} for electron transport,
$\bm{J}(t)=\sigma \bm{E}[1-\exp{(-t/\tau)}]$, where $\sigma=e^2n_e\tau/m_e$ is conductivity and $\tau$ is the typical collision time of background electrons, which is usually much smaller than $2\pi/\omega_{pe}$.
In the ``layered density'' method, 
we have $\bm{J}_{e1}(t)\sim en_{e1}\bar{\bm{v}}_{e1}\sim e^2n_{e1}\bm{E}\delta t/m_e\sim\sigma\bm{E}[1-\exp{(-\delta t/\tau)}]$. 
Within one time step, if the product of $n_{e1}\delta t$ is much larger than $n_e\tau$, then one can not distinguish the differences whether all the background electrons $n_e$ or just $n_{e1}$ involve in Ohmic return current or/and resistive electromagnetic fields calculations. 
After the abrupt building of Ohmic return current, whose following evolution is much slow, 
any small variations of Ohmic return current $\delta \bm{J}(\delta t)$ can be compensated 
by the redistribution of $n_{e1}$ and $\bar{\bm{v}}_{e1}$ within $\delta t$.  
Here we present an empirical formula, where for the given PIC simulation time step $\delta t$ 
and \textit{initial} background plasma temperatures $T_e$, the threshold density of ele-$1$ is 
\begin{equation}
\label{threshold}
n^{\text{th}}_{e1}\sim10^{19}\times T^{3/2}_e\ [\text{eV}]/\delta t\ [\text{fs}]\ \text{cm}^{-3}.
\end{equation}
In the simulation set-up, we have $n_{e1}\gg n^{\text{th}}_{e1}$. Note the ``layered density'' method is still an empirical method instead of an rigorous numerical scheme. To ensure that the simulation results are physically correct, we would suggest to re-run the simulation by increasing the ele-$1$ density twice to confirm the convergence of finial results. 

Similarly, Sentoku and Kemp proposed a ``reduced'' PIC method \cite{pic_collision3} to artificially reduce the plasma density in collisional PIC simulation 
when it exceeds an upper-limit value, e.g., $100n_c$. 
In this approach a macro-particle has two weights: one is its real weight which is used to calculate Coulomb collision
and another is a reduced weight to calculate the current for the field solver. 
However to use this approach, one should make sure that the numerical noise at high-density region is controlled to be quite low. 
Because the energy of a macro-particle gaining from the noise can be easily amplified 
by the ratio of the real weight to the reduced one. With ``reduced'' PIC method, 
Chrisman et al. \cite{reducepic} 
have performed a group of integrated fast ignition simulations with the core density as high as $20000n_c$ or $100$ g$/$cm$^3$. 

\section{Applications}

In this part, the electron heating/acceleration at relativistically intense laser-solid interactions in the presence of large scale pre-formed plasmas \cite{ta1,ta2} is re-investigated by LAPINE code (Appendix B), which could include almost ``all'' the coupled physical mechanisms. Thanks to the ``layered density'' method and the coupled high-order numerical scheme and current smoothing technique, the simulation grid size can be significantly larger than the plasma Debye length. Larger simulation grid would dramatically reduce the simulation burden, which makes it possible for a small cluster with only $10$ nodes (with each node containing $12$ cores) to simulate realistic laser-solid interactions in large scales, both specially and temporally.  

\subsection{1D simulations}

\begin{figure}
\includegraphics[width=8.50cm]{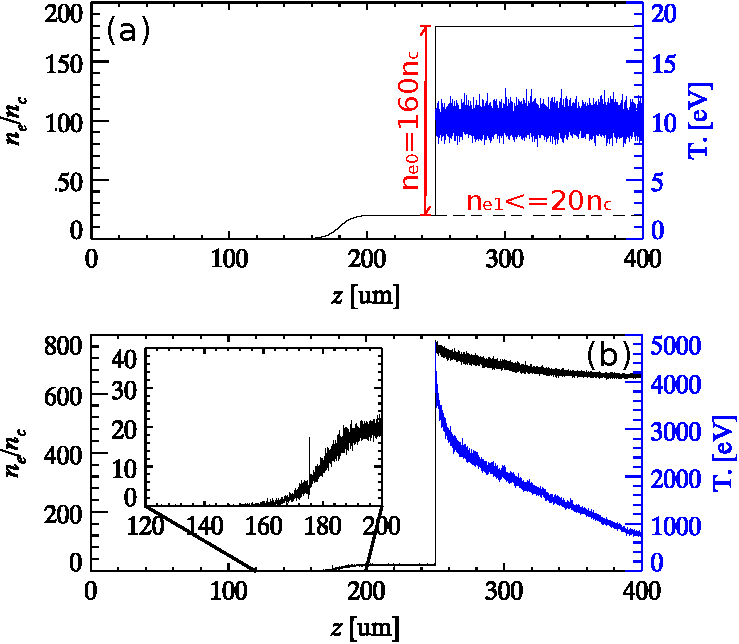}
\caption{\label{fig10} (color online) (a) The initial parameter set-up, with pre-plasma scale-length $10\ \mu\text{m}$, initial density $180n_c$ ($Z=3$) and temperature $10\ \text{eV}$. In the ``layered density'' method, density of ele-$0$ is $n_{e0}=160n_c$ and ele-$1$ is $n_{e1}=20n_c$. 
Here $n_c=1.1\times10^{21}\ /\text{cm}^3$ is the corresponding critical density of electromagnetic wave with wavelength $1\ \mu\text{m}$.  
(b) Electron density and temperature at the end of simulations.}
\end{figure}   

\begin{figure}
\includegraphics[width=8.50cm]{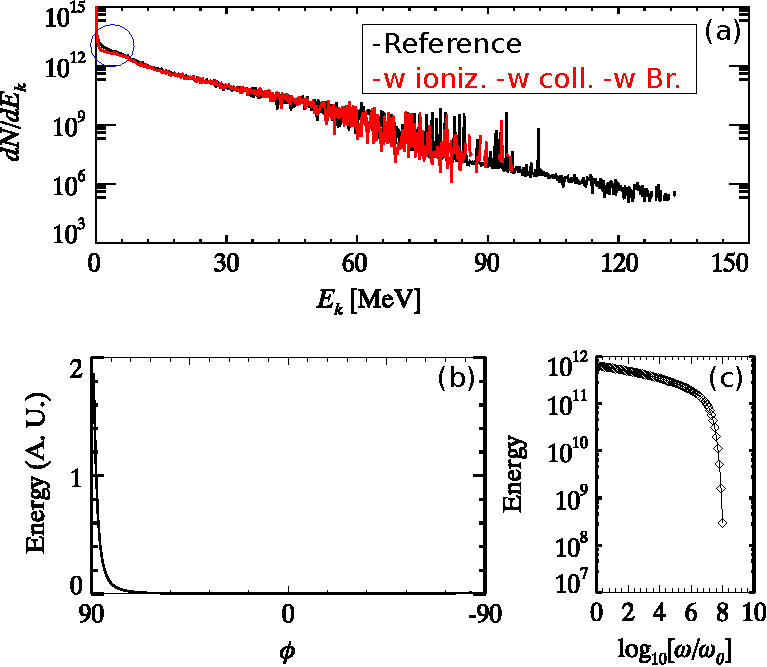}
\caption{\label{fig11} (color online) (a) The finial energy spectra of electrons, with black-line represents the reference case without considering atomic processes, and red-line represents the one including both ionization and collision with Bremsstrahlung radiation corrections. (b) The angular distribution of emitted photons. (c) The frequency spectra of emitted photons, where we have plotted $\int_{\hbar\omega_{\text{k}}}^{\infty}[dE/d(\hbar\omega)]d(\hbar\omega)$ as function of cut-off frequency $\omega_{\text{k}}$. 
Note $\hbar\omega_{0}=1.24$ eV, corresponding to the energy of a photon with wavelength $1\ \mu\text{m}$.}
\end{figure}

\begin{figure*}
\includegraphics[width=16.50cm]{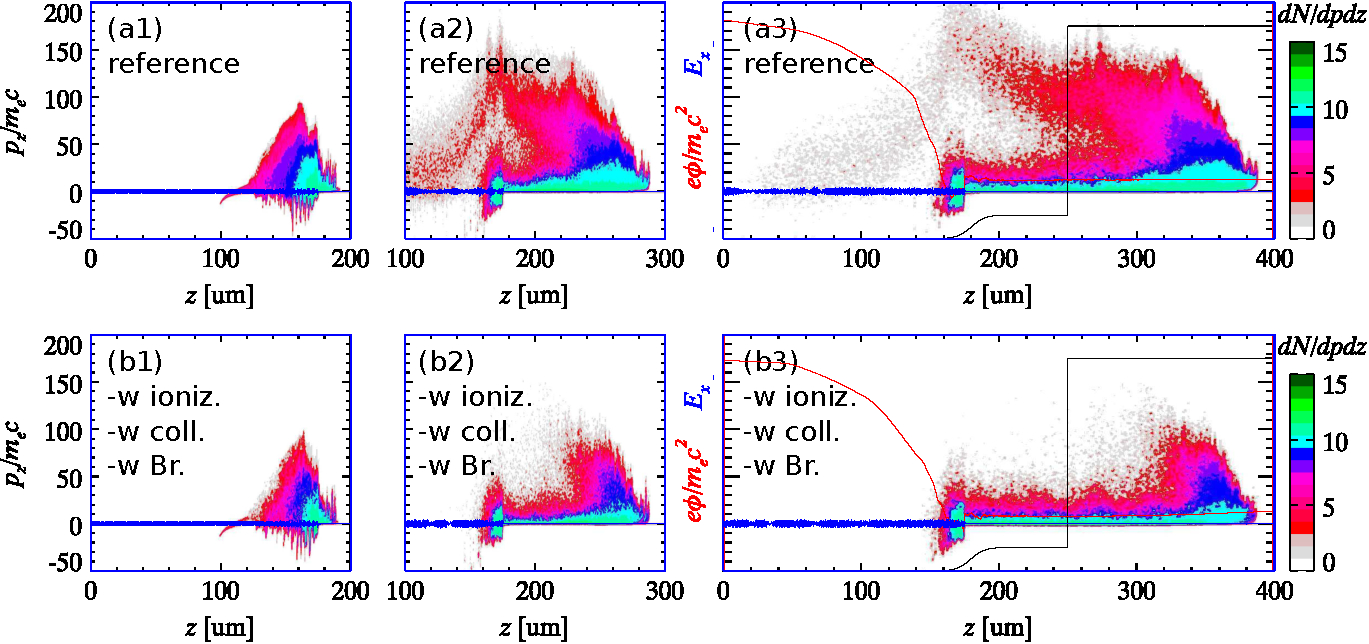}
\caption{\label{fig12} (color online) The $z$-$p_z$ phase space plot of electrons, with the same simulation parameters as shown in Fig.\ \ref{fig10}. (a) The one without considering ionization, collision and Bremsstrahlung radiation correction. (b) The one turning on both ionization and collision with Bremsstrahlung radiation corrections. 
Different columns represent values at different times, 
here $t=0.67$ ps for (1), $t=1.0$ ps for (2) and $t=1.3$ ps for (3). 
The red-curves covered on the phase-space plots are the electrostatic potential curves ($\int^{z}E_z dz$), normalized by $-e\phi/m_ec^2$.
The blue lines are the $E_x(\times0.25)$ components of the superposition of incoming and reflected laser pulses.}
\end{figure*}

\begin{figure*}
\includegraphics[width=16.50cm]{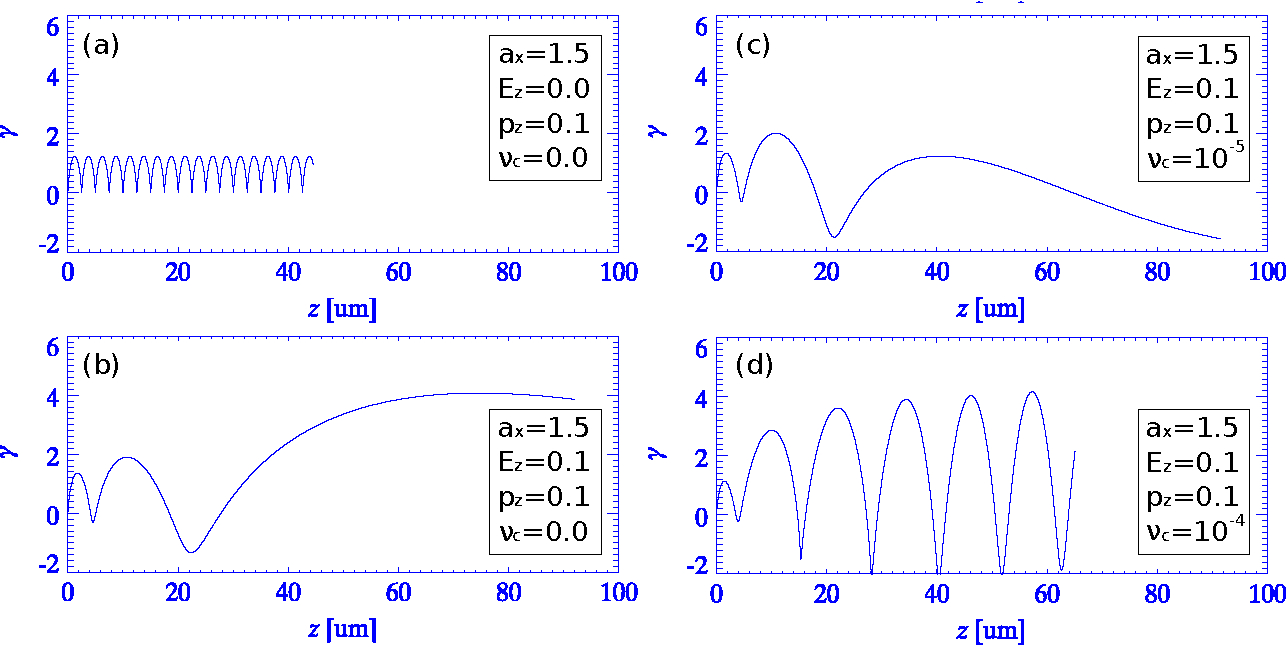}
\caption{\label{fig13} (color online) Dynamics of an electron calculated with single-particle-simulations. The gained energy from laser beam as function of propagation length. The total simulation time is $100T_0$.
(a) An electron with initial momentum $p_z=0.1$, a single laser pulse of amplitude $a_x=1.5$.
(b) An electron with initial momentum $p_z=0.1$, a single laser pulse of amplitude $a_x=1.5$, 
and a constant external electric field of $E_z=-0.1$.
(c) An electron with initial momentum $p_z=0.1$, a single laser pulse of amplitude $a_x=1.5$, 
a constant external electric field of $E_z=-0.1$ and initial collision frequency of $10^{-5}$.
(d) An electron with initial momentum $p_z=0.1$, a single laser pulse of amplitude $a_x=1.5$, 
a constant external electric field of $E_z=-0.1$ and initial collision frequency of $10^{-4}$.}
\end{figure*}

\begin{figure}
\includegraphics[width=8.50cm]{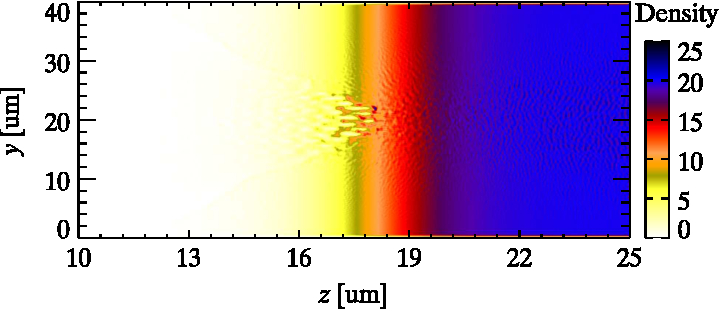}
\caption{\label{fig14} (color online) Results of 2D PIC simulations. 
The plasma density perturbations in front of the target at the end of simulation time.}
\end{figure}

\begin{figure}
\includegraphics[width=8.50cm]{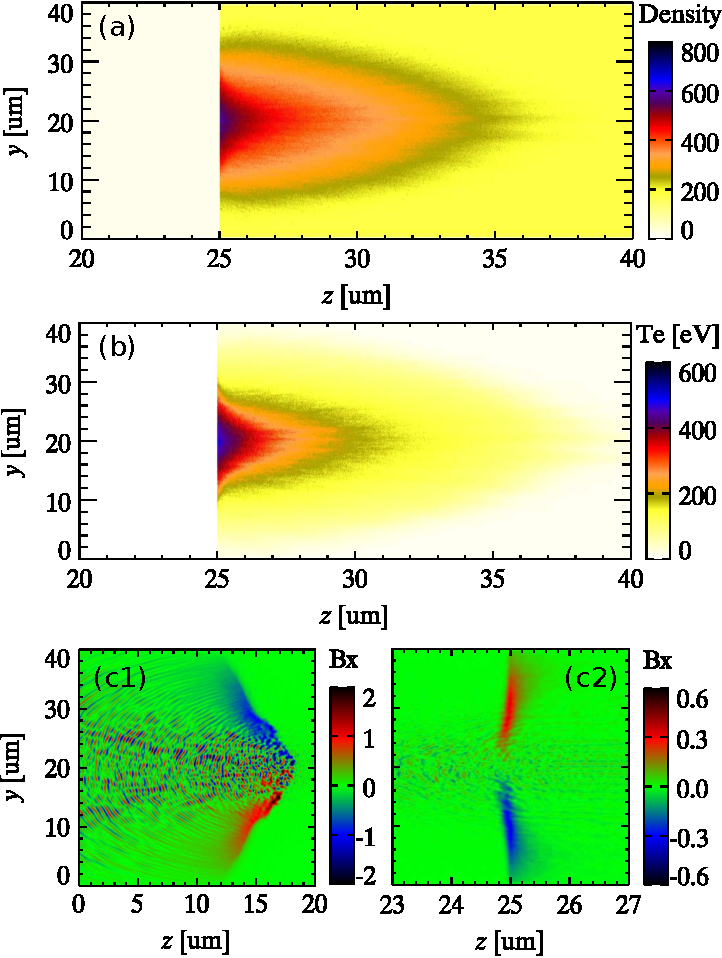}
\caption{\label{fig15} (color online) Results of 2D PIC simulations. 
(a) The plasma density in the inner part of the target at the end of simulation time.
(b) The plasma temperature in the inner part of the target at the end of simulation time.
(c1) The magnetic fields generated by the forward propagating fast electrons.
(c2) The resistive magnetic fields generated by the Ohmic return current.}
\end{figure}

\begin{figure}
\includegraphics[width=8.50cm]{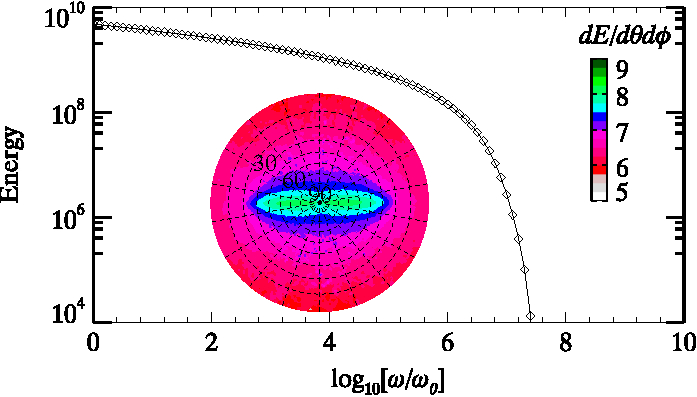}
\caption{\label{fig16} (color online) Results of 2D PIC simulations. 
(a) The angular distribution of emitted photons. (b) The frequency spectra of emitted photons, where we have plotted $\int_{\hbar\omega_{\text{k}}}^{\infty}[dE/d(\hbar\omega)]d(\hbar\omega)$ as function of cut-off frequency $\omega_{\text{k}}$. 
Note $\hbar\omega_{0}=1.24$ eV, corresponding to the energy of a photon with wavelength $1\ \mu\text{m}$.}
\end{figure}

The simulation set-up of 1D PIC is shown in Fig.\ \ref{fig10} (a). The simulation box is $400\ \mu\text{m}$, 
an Al target with maximal density of $2.7\ \text{g}/\text{cm}^3$ (or ion density of $60n_c$ for laser of wavelength $1\ \mu\text{m}$) and 
temperature of $10\ \text{eV}$ is applied.  
The simulation grid size is $\delta z=0.02\ \mu\text{m}$, 
which is smaller than the skin depth $l\sim0.021\ \mu\text{m}$. 
In the ``layered density'' method, density of ele-$1$ is $n_{e1}(z)=n_{e1}/(1+\exp[-2(z-180)/L_p])$, where $n_{e1}=20n_c$ is the solid plasma density and $L_p=10$ is the pre-plasma scale-length. The density of ele-$0$ is $n_{e0}(z)=160n_c$ when $z>250$ and otherwise $n_{e0}(z)=0$. 
The simulation time step is $\delta t=3.9\times10^{-2}\ \text{fs}$. Therefore, according to Eq.\ (\ref{threshold}),
the corresponding density threshold is $n^{\text{th}}_{e1}=7n_c$, which is much smaller than $n_{e1}=20n_c$. The laser intensity is $10^{20}\ \text{W}/\text{cm}^2$ or normalized amplitude $a=8.54$ (with laser wavelength $1\ \mu\text{m}$). It enters the simulation box from the left boundary,
where the laser amplitude rises over $33$ fs to $a=8.54$ and then remains constant. 

The finial electron density and temperature (at $t=1.3\ \text{ps}$) are presented in Fig.\ \ref{fig10} (b). We can see along the laser propagation direction, both electron density and temperature decrease rapidly. As the thermal equilibrium is not yet established within such a short time, here we use ``temperature'' to represent the average kinetic energy of electrons. In Fig.\ \ref{fig10} (b), at $z=380$, temperature is $T_e=1000\ \text{eV}$, while the corresponding electron density is $660n_c$ (or $Z=11$).
This is already much smaller than the thermal equilibrium ionization degree, $Z=12.9$ with $T_e=1000\ \text{eV}$ as shown in Fig.\ \ref{fig1} (b). 
At earlier times, as expected, the departure from the thermal equilibrium values could be more significant than at finial times. 
The detailed comparison is not shown in this paper, but one can refer to Fig.\ \ref{fig1} (a) to see the evolution of ionization dynamics with time.
 
In Fig.\ \ref{fig11} (a), we have presented the electron energy spectra, comparing different cases without/with ionization, 
collision and Bremsstrahlung radiation interactions. The black-line represents the reference case without considering these atomic processes, 
and red-line represents the one including these atomic processes. We can see that the electron ``cut-off'' energy is significantly lowered by $25\%$. 
In addition, as the blue circle shows, the number of electrons with low energies, i.e., less than $3$ MeV, is also significantly reduced. The latter one can be interpreted by collisional damping. While for the former one, it might due to the Bremsstrahlung radiation, as this radiation is very efficient for those energetic electrons, with energy larger than $10$ MeV.
The angular distribution of emitted photons is shown in Fig.\ \ref{fig11} (b). We can see that the direction of emitted photons is along the laser propagation direction, with a small diffraction angle of $\delta \phi\sim10$ degree.  
The frequency spectra of emitted photons, where we have plotted $\int_{\hbar\omega_{\text{k}}}^{\infty}[dE/d(\hbar\omega)]d(\hbar\omega)$ as function of cut-off frequency $\omega_{\text{k}}$, is shown in Fig.\ \ref{fig11} (c).
The cut-off frequency is of 
$\omega_{\text{k}}\sim10^8\omega_0\sim100$ MeV, which is equal to the ``cut-off'' energy of electrons. 

However the Bremsstrahlung radiation alone can not fully explain the $25\%$ reduction of ``cut-off'' energy. 
This is because, as shown in Fig.\ \ref{fig3} (c), for Al, 
the stopping power of electrons with energy $100$ MeV is only $5\times10^{-3}$ MeV$/\mu$m. 
For a propagation distance of $200$ $\mu$m, the energy reduction is only $1$ MeV. As the values of stopping power heavily depend on target materials, 
$S\sim A^2$ ($A=29$ for Cu and $A=79$ for gold), 
if the target material is of Cu and/or gold, the energy reduction can be as high as $5$ MeV and/or $37$ MeV. 
Note the material dependence of electron heating/acceleration is not the purpose of this paper, which shall be detailed in the following works.
The present paper is focused on presenting a global simulation framework and addressing the influences of the coupled atomic processes on laser solid interactions. For the considered Al target, Bremsstrahlung radiation only contribute $\sim1$ MeV energy reduction, therefore, there must exist other mechanisms that significantly reduce electron heating/acceleration.    

In order to figure out the other mechanisms that significantly reduce electron heating/acceleration, 
we now refer to the phase-space plots of electrons, as shown in Fig.\ \ref{fig12}. 
The phase-space density $dN/dzdp$ gives a value proportional to the number of electrons found between $z$ and $z+dz$ having longitudinal momentum ranged between $p_z$ and $p_z+dp_z$. Energetic electrons are generated in front of the target. 
Fig.\ \ref{fig12} (a) and (b) show the dynamics of electron heating/acceleration at $t=0.67$ ps and $t=1.0$ ps respectively. 
Fig.\ \ref{fig12} (c) shows the global pictures containing both electron heating/acceleration and transportation at $t=1.3$ ps.   
We can see that electron heating/acceleration is dramatically depressed when collision is included in front of the target. As we know, in front of the target, plasma density therein is low, therefore collision effect is relatively small when compared with electromagnetic effects. In the following, we shall explain the reason, even though the collision damping in front of the target is week, it can still have significant effects on electron heating/acceleration.

When a laser propagates in under-dense preformed plasma, 
part of electrons are swept away in the forward direction by the laser ponderomotive force, leaving behind immobile ions. 
The electric field $E_z$ due to charge separation within the under-dense plasma region tries to pull the electrons in the backward direction. 
When the laser arrives at the critical density surface and is reflected back, the ponderomotive force of the reflected laser pulse can further accelerate the electrons in the backward direction. The synergetic effects by this longitudinal charge separation field $E_z$ and the ponderomotive force of the reflected laser pulse can efficiently accelerate electrons in the backward direction. 
This backward acceleration is clearly figured out in Fig.\ \ref{fig12} (a) and (b).
In fact, when the incident laser arrives at the critical density surface and is reflected back, due to the formation of the steep interface of electron density, a strong delta-like charge separation field or the step-like electrostatic potential, as shown in Fig.\ \ref{fig12}, is build up therein.
This field is strong enough to drive electrons to very high velocity within very short time and short length. This ``initial large velocity'' can significantly simplify 
our following analysis. Imagine we are standing on the frame of a backward propagating electron, we will find that the incident laser pulse is oscillating very fast, and its only contribution to the motions of the electron is to increase its mass by a factor $\gamma=(1+a^2/2)^{1/2}$ in an average way, however the reflected laser pulse is oscillating so slow that this electron can be captured and continually be accelerated backward by its ponderomotive force. 
 
From Woodward-Lawson theorem, we know that a single electron in vacuum, oscillating coherently with a propagating plane laser pulse would gain zero cycle averaged energy since the electron energy gain in one half cycle is exactly equal to the energy loss in the next half cycle. In Fig.\ \ref{fig13}, we have presented single particle simulations. It shows the dynamics of an electron of initial momentum $p_z=0.1$ under a laser pulse of amplitude $a=1.5$. The maximal energy gain from laser field is $m_ec^2a^2/2=1.125$, and this value is the same as obtained by single particle simulations, Fig.\ \ref{fig13} (a).
However, when there exists an external electric field, 
even though this field is very week, the Woodward-Lawson theorem can be broken and the electron can obtain non-zero energy from the synergetic effects by the external electric field and the laser pulse. If the extension of external electric field is of infinity, the electron will always stay in phase with the laser and be accelerated to energy of infinity. In Fig.\ \ref{fig13} (b), when we add a small external magnetic field, $E_z=-0.1$, the electron dynamics is dramatically changed. The energy gain is significantly higher than $m_ec^2a^2/2=1.125$. In a recent work \cite{ta1}, we have proved that the maximal energy gain is scaled as $\sim a L^{1/2}$, where $L$ is the propagation length. 

In front of the target, although the charge separation is very small, it has significant influences on electron dynamics. Similarly, the week collisional damping might also play important roles in this interactions. Let us firstly estimate the collision frequency. The considered plasma is of $1.0n_c$, temperature is of $\gamma m_ec^2$, the collision frequency is $\nu_{\text{c}}=10^{-5}\gamma^{-3/2}$. In the single particle simulation, we have add this week collisional damping term $-\nu_{\text{c}}\bm{p}_e$ into the electrons' Equation of Motion. As shown in Fig.\ \ref{fig13} (c), 
when the collision frequency is $10^{-5}\gamma^{-3/2}$, the maximal energy gain within $100T_0$ is $2.0$. In Fig.\ \ref{fig13} (d), when the collision frequency is $10^{-4}\gamma^{-3/2}$, the maximal energy gain within $100T_0$ is $4.0$. 
Although the energy gains of electrons are significantly depressed when compared with collision-less cases, they are still much larger than that value $1.125$.

\subsection{2D simulations}

In this section, we shall present how the ``layered density'' method PIC works in 2D simulations. Here to avoid the extensive calculation burden, 
we use a smaller simulation box and shorter laser pulse duration. The simulation box is $40\ \mu\text{m}\times40\ \mu\text{m}$ 
($L_z\times L_y$), with grid size $\delta z=0.02\ \mu\text{m}$ and $\delta y=0.1\ \mu\text{m}$. 
The pre-plasma scale-length used in 2D simulation is $5\ \mu\text{m}$. 
Other parameters, like plasma density division, temperatures and laser intensity, are the same with 1D simulation.

The plasma distortion in front of the target is shown in Fig.\ \ref{fig14}. In this region, energetic electron are generated directly by laser fields. When these electrons propagate into the bulk solid, abundant plasma and atomic interactions take place therein. As shown in Fig.\ \ref{fig15} (a) and (b), 
collision ionization would dramatically increase the electron density. Typically, the ionization and plasma temperature decrease rapidly along the electron propagation direction. When compared with 1D simulations, we find some filamentation structures in the electron density and temperature distributions. This kind of filamentation is due to two-stream or/and Weibel instabilities.  

The propagation of electrons could generate magnetic fields. Except electromagnetic instabilities, like Weible instability, 
there are two sources that can generate strong magnetic fields: i) $\nabla\times\bm{B}=4\pi\bm{J}/c$; ii) $-\partial \bm{B}/\partial t=\nabla\times\bm{E}$.
As shown in Fig.\ \ref{fig14} (c1), in the front of target, this magnetic field is fully due to the forward $\bm{J}_{e}$, i.e., the i) generation mechanism, which could cause divergence of electron beams. When these electrons propagate into solid, the forward $\bm{J}_{e}$ is quickly neutralized by Ohmic return current. The total current density is close to zero, therefore the former mechanism is not effective any more. However, because of the strong collision effect, a resistive electric field $E_z=\bm{J}_{e}/\sigma$ can be generated, which in turn could produce the resistive magnetic field, through the ii) generation mechanism. This magnetic field, as shown in Fig.\ \ref{fig14} (c2), could collimate the electron beam.     

In Fig.\ \ref{fig16}, we also present the angular distribution (a) and energy spectra (b) of emitted photons. The diffraction angle obtained in 2D simulation is significantly higher than 1D, which can be as large as $30$ degree. The ``cut-off'' frequency of emitted photons is significantly smaller than 1D simulations. This is because, short laser pulse and pre-plasma scale-length are used in 2D simulations. The maximal electron energy in 2D simulations is much smaller than that in 1D simulations.  



\section{Conclusions and discussions}

To summary, we have presented a full PIC framework, which enables us to calculate intense laser-solid interactions in a ``first principle'' way, covering almost ``all'' the coupled physical mechanisms. For ionizations, we have taken into account CI, RE and IPD. For collisions, we have taken into account both bound and free electrons' contributions. A modified Coulomb logarithm is used in the binary collision model, which has the ability to deal with collisions at low temperatures, when the closest approach distance is larger than Debye length. For energetic electron-atom/ion collisions, Bremsstrahlung radiation correction is also included in our model.

The ``layered density'' method PIC is proposed to simulation plasma dynamics at extremely high densities. 
The numerical self-heating of PIC simulations with solid-density plasmas can be well controlled by this method. 

Especially, the electron heating/acceleration at relativistically intense laser-solid interactions in the presence of large scale pre-formed plasmas is re-investigated by this PIC code. Results indicate that collisional damping (even though it is very week) 
can significantly influence the electron heating/acceleration in front of the target. 
Furthermore the Bremsstrahlung radiation will be enhanced by $2\sim3$ times when the solid is dramatically heated and ionized. 
For the considered laser of intensity $10^{20}\ \text{W}/\text{cm}^2$ and solid Al target with pre-plasmas scale-length $10\ \mu\text{m}$, collision damping coupled with ionization dynamics and Bremsstrahlung radiations is shown to lower the ``cut-off'' electron energy by $25\%$. In addition, the resistive electromagnetic fields due to Ohmic-heating also play a non-ignorable role and must be included for realistic laser solid interactions. 

\begin{acknowledgments}    
D. Wu wishes to acknowledge the financial support from German Academic Exchange Service (DAAD) and China Scholarship Council (CSC). This work was also supported by National Natural Science Foundation of China (No. 11605269, 11674341 and 11675245).
\end{acknowledgments}

\begin{appendix}
\section{The calculation of Coulomb logarithm}
To calculate Coulomb logarithm, one of the practical approaches, as used in Takizuka, Nanbu and Sentoku's models, is to sum binary collisions over a distance of the order of the Debye length. Under the potential of $1/r$, the differential cross-section reads, $\sigma(\theta)\sim1/\sin^4(\theta/2)$, and the Coulomb logarithm reads, $L\sim\int_0^{\pi}\sin{\theta}\sin^2(\theta/2)\sigma(\theta)d\theta\sim\ln(\sin(\theta/2))|_0^{\pi}$. This integration is not a convergent value, when $\theta\rightarrow0$. 
While in plasmas, the potential of a charged particle should be screened. 
When $b$ (i.e., the distance of closest approach between the two charges) is larger than $\lambda_{\text{D}}$, the potential is artificially set to be zero. Therefore, the lower limit $\theta_{\min}$ of scattering angle is obtained when $b=\lambda_{\text{D}}$, i.e., $\theta_{\text{min}}/2=b/\lambda_{\text{D}}$. Thus we have $L\sim\ln(\lambda_D/b)$. 
  
However instead of the above method, a rigorous way is to sum full binary collisions with all particles using the screened potential 
$\exp(-r/\lambda_{\text{D}})/r$. Acted by this screened potential, the differential cross-section reads, 
$\sigma(\theta)\sim1/(\sin^2(\theta/2)+\eta)$, where $\eta$ is the smallest value between $\hbar/\gamma\beta\lambda_{\text{D}}$ (quantum) and $Z_a Z_b e^2/m_e\beta^2\lambda_{\text{D}}$ (classical). The Coulomb logarithm $L\sim\int_0^{\pi}\sin{\theta}\sin^2(\theta/2)\sigma(\theta)d\theta$ by applying the new differential cross-section is $L\sim\ln[1+2\eta-\cos(\theta)]|_0^{\pi}$. This is a convergent value, with $L\sim\ln[(1+\eta)/\eta]$. This expression of Coulomb logarithm will converge to $L=\ln{(\lambda_{\text{D}}/b)}$ when $b\ll\lambda_{\text{D}}$ for high temperature plasmas.

\section{A brief introduction to LAPINE code}

LAPINE \cite{lapine} is the abbreviation of $\underline{\text{LA}}$ser-$\underline{\text{P}}$lasma-$\underline{\text{INE}}$raction. 
It is one of the first-generation PIC codes fully developed by Chinese. LAPINE is a parallel PIC code, written in C++ language, 
capable of performing both 1D and 2D$/$3D simulations. Both 1D and 2D/3D versions are self-consistently written into a single group of files. 
Set-up of 1D or 2D/3D is defined in pre-compilation to compile the code into the specific LAPINE-1D or LAPINE-2D/3D. 

\textit{Physical models}--Many advanced physical modules have been implemented into LAPINE code, which include 
bulk ionization \cite{hedp3} (coupling impact ionization, electron-ion recombination and ionization potential depression by surrounding plasmas), binary collisions \cite{hedp4} (partially ionized plasmas and also pure plasmas for all temperature ranges),
field ionization \cite{fieldionization} and quantum electrodynamics \cite{qed} modules. Note all the physical modules have been well benchmarked and applied for related physical research.  

\textit{Numerical scheme}--High order Electromagnetic field solver, high-order-particle-shape and current-smooth-technique have been implemented into LAPINE to improve its ability calculating high density plasmas. The proposed ``layered density'' method is firstly implemented into the LAPINE code, showing strong power in performing laser-solid simulations in large scale. 
\end{appendix}

{}

\end{document}